\documentclass[reprint,notitlepage,amsmath,amssymb,jmp,superscriptaddress,onecolumn]{revtex4-2}
\usepackage{amsmath}
\usepackage{graphicx}
\usepackage{placeins}
\usepackage{xspace}
\usepackage[table,xcdraw]{xcolor}
\usepackage{booktabs}
\usepackage{tabularx}
\usepackage[acronym]{glossaries}

\newcolumntype{Y}{>{\centering\arraybackslash}X}
\newcommand{\coo}{$\text{CO}_{2}$\xspace}
\newacronym[\glslongpluralkey={Metal Organic Frameworks}, \glsshortpluralkey={MOFs}]{mof}{MOF}{Metal Organic Frameworks}
\newacronym{dft}{DFT}{Density Functional Theory}
\newacronym{dmet}{DMET}{Density Matrix Embedding Theory}
\newacronym{hf}{HF}{Hartree Fock}
\newacronym{dodbc}{dodbc}{2,5-dihydroxy-1,4-benzenedicarboxylic acid}
\newacronym{m06l}{M06L}{Minnesota 2006 Local}
\newacronym{sqd}{SQD}{Sample-Based Quantum Diagonalisation}

\begin{document}
\title{A Multi-Scale Quantum Framework for Evaluating Metal-Organic Frameworks in Carbon Capture}
\date{\today}

\author{Tom W. A. Montgomery}
\affiliation{Capgemini Quantum Lab, Capgemini, 147-151 Quai du Président Roosevelt,92130 - Ile-de-France, Issy-les-Moulineaux, France}
\author{Adrian Varela-Alvarez}
\affiliation{Capgemini Quantum Lab, Capgemini, 147-151 Quai du Président Roosevelt,92130 - Ile-de-France, Issy-les-Moulineaux, France}
\author{Sam Genway}
\affiliation{Capgemini Quantum Lab, Capgemini, 147-151 Quai du Président Roosevelt,92130 - Ile-de-France, Issy-les-Moulineaux, France}
\author{Philip Llewellyn}
\affiliation{TotalEnergies, Tour Coupole, 2, Place Jean Millier, Paris, Ile-de-France, 92078}
\author{Phalgun Lolur*}
\affiliation{Capgemini Quantum Lab, Capgemini, 147-151 Quai du Président Roosevelt,92130 - Ile-de-France, Issy-les-Moulineaux, France}

\begin{abstract}
\acrfullpl{mof} are promising materials to help mitigate the effects of global
warming by selectively absorbing \coo for direct capture. Accurate quantum chemistry simulations are
a useful tool to help select and design optimal \acrshort{mof} structures, replacing costly or impractical
experiments or providing chemically inspired features for data-driven approaches such as
machine learning. However, applying simulations over large datasets requires efficient simulation methods 
such as \acrfull{dft} which, despite often being accurate, introduces uncontrolled approximations and a
lack of systematic improvability. In this work we outline a hierarchical cluster model that includes a
recently developed quantum embedding that provides a more systematic approach to efficiently tune accuracy. We apply this workflow to calculate the binding affinity for a small set of \acrshort{mof} structures and
\coo using experimentally measured heat of adsorption as a reference. Since quantum embeddings have also
been proposed as a framework to accelerate the utility of quantum hardware, we discuss some of the benefits
and challenges of integrating quantum solvers into the workflow outlined in this work.
\end{abstract}

\maketitle
\par{$^*$ Corresponding author - phalgun.lolur@capgemini.com}
\section{Introduction}
\label{sec:intro}

Carbon dioxide (\coo) has emerged as the primary driver of human-induced climate change, with atmospheric concentrations reaching unprecedented levels in recent decades \cite{mikhaylov2020a}. The Intergovernmental Panel on Climate Change (IPCC) has emphasized that limiting global temperature rise to 1.5°C above pre-industrial levels requires not only dramatic reductions in \coo emissions but also the implementation of negative emission technologies \cite{bui2018a}. This challenge has sparked intensive research into various carbon mitigation strategies, as no single approach can adequately address the scale of the problem. Carbon capture, which involves the removal of \coo from either point sources or directly from the atmosphere for subsequent storage or utilization, represents one of the most promising technological solutions \cite{bui2018a,osman2020a}. The captured \coo can be either stored in geological formations (Carbon Capture and Storage, CCS) or converted into valuable products (Carbon Capture and Utilization, CCU), providing both environmental and potential economic benefits.

Metal-organic frameworks (MOFs) have emerged as particularly promising materials for carbon capture applications, offering exceptional versatility and tunable properties \cite{osman2020a,and2022a}. These crystalline materials consist of metal ions or clusters (nodes) connected by organic ligands (linkers) to form three-dimensional networks with well-defined pore structures. The remarkable feature of MOFs lies in their unprecedented porosity, with some materials exhibiting specific surface areas exceeding 7000 m²/g, far surpassing traditional porous materials such as zeolites and activated carbons \cite{choe2021a}. Their modular nature allows for precise control over pore size, shape, and chemical environment, making them highly adaptable for specific applications \cite{keith2021a, pollice2021a}.

In terms of their composition, MOFs exhibit a deceptive chemical simplicity, typically are made of one or few metal centres and one or two organic linkers. However, this apparent simplicity belies the complexity of their structure-property relationships. Even minor modifications to either the metal center or the organic linker can induce profound changes in the framework's structure, stability, and adsorption properties \cite{kim2021a}. For instance, the presence of defects such as missing linkers or the substitution of one metal ion for another \textit{while maintaining the same organic linker} can dramatically alter the framework's binding affinity for CO2, pore accessibility, and overall capture performance. These subtle variations create both opportunities and challenges: while they offer vast possibilities for material optimization, they also make the prediction of MOF properties particularly challenging. The complex interplay between electronic structure, geometric arrangements, and guest-host interactions requires sophisticated computational approaches to accurately model and predict their behavior \cite{choe2021a}.

The development of MOFs for carbon capture applications has seen remarkable progress, with thousands of structures synthesized and characterized to date. However, this represents only a tiny fraction of the theoretically possible combinations of metals and organic linkers. The vast chemical space available for MOF design, combined with the sensitivity of their properties to small structural changes, has led to increased interest in computational screening and machine learning approaches to accelerate the discovery of optimal materials for CO2 capture. Understanding and accurately predicting the relationship between MOF composition, structure, and CO2 capture performance has therefore become a critical challenge in materials science, requiring advanced computational methods that can balance accuracy with computational efficiency.

Simulations can replace many costly or impractical experiments or provide chemically inspired machine learning features for data-driven approaches. However, the number of simulations scales with the size of the dataset used to train the model, which often needs to be large to produce useful predictive or generative machine learning models. Density Functional Theory (DFT) has emerged as one of the most widely used computational methods in materials science and chemistry since its theoretical foundations were established by Hohenberg, Kohn, and Sham in the 1960s. By reformulating the many-body electronic structure problem in terms of the electron density rather than the many-electron wavefunction, DFT provides a computationally tractable approach to predict material properties from first principles. Due to its efficiency, DFT is an attractive approach, especially for the larger systems involved in materials science. While DFT can potentially be highly accurate, in practice DFT's accuracy often depends on the specific exchange-correlation functional, and there are no clear guiding principles for selecting the right functional for the right problem \cite{cramer2004a, jensen2017a}. Furthermore, even if one selects a functional that produces satisfactory accuracy, there is no systematic way of improving its performance if higher accuracy is required.

The limitations of DFT become particularly apparent when dealing with strongly correlated systems or when precise energetics are required for chemical reactions. Different functionals may perform well for certain properties but fail for others, making it challenging to establish a universal approach for materials discovery and characterization.

Quantum embedding methods offer an alternative route by fragmenting large quantum systems into multiple smaller auxiliary "cluster" problems, providing a way to use ab initio correlated methods in a computationally feasible manner \cite{RevModPhys.78.865, PhysRevLett.109.186404}. These methods combine the efficiency of treating the majority of the system at a lower level of theory while maintaining high accuracy for the chemically relevant regions. The key advantage lies in their ability to scale favorably with system size while potentially maintaining the accuracy of high-level quantum chemical methods for the most important regions of the system.

Booth and co-workers have introduced the wavefunction-based embedding (EWF) \cite{nusspickel2022a}, which builds on DMET with the ability to systematically converge properties of real materials controlled by a single, rapidly convergent parameter. This makes it an attractive method when trying to balance efficiency and accuracy, which is key when applying such methods across large datasets of materials. The EWF approach offers significant advantages over traditional embedding methods through its systematic improvability via a well-defined convergence parameter, reduced computational complexity compared to full system treatments, and maintained accuracy for strong correlation effects.

In this work, we outline a workflow to incorporate this systematically improvable embedding to calculate the binding energy between MOFs and \coo. To validate this method, we apply it to the small set of MOF structures listed in Table \ref{table:mofs}, comparing it to the measured heat of adsorption shown in the same table. This allows us to perform a small study in which the quantum embedding method is replaced by HF and DFT methods with several different functionals. We outline a scheme in which the quantum embedding scheme performs comparatively to the best performing DFT method herein employed (M06L) and therefore a candidate to expand to a larger scale study over a wider range of MOF structures.

Quantum embeddings have been seen to bring the benefits of groundstate solvers that use quantum computers to larger chemical systems \cite{ALEXEEV2024666}. We focus on the fact that the results in this work are based on approach direct projection of wave function amplitudes, rather than density matrices. In general, accessing wave function amplitudes from states represented as a quantum circuit is difficult, so it is useful to outline a strategy for how to best combine the emerging benefits of quantum hardware with these types of embedding methods. This consideration becomes particularly relevant as quantum computing hardware continues to advance and may soon be capable of handling increasingly complex chemical systems.

The integration of quantum embedding methods with quantum computing represents a significant frontier in computational chemistry. The potential for quantum advantage in treating strongly correlated subsystems must be balanced against the technical challenges of extracting wave function information from quantum states. Furthermore, the development of hybrid classical-quantum algorithms for materials simulation presents an opportunity to leverage the strengths of both computational paradigms. This synthesis of quantum embedding with quantum computing capabilities may ultimately provide a path toward more accurate and efficient materials modeling, particularly for systems where traditional electronic structure methods struggle to provide reliable results. 

While computational methods have generated thousands of hypothetical MOF structures in recent years, it's important to acknowledge that the successful experimental realization of computationally-predicted MOFs remains challenging. To date, NU-100, developed by Farha and colleagues, stands as one of the rare examples where computational design directly led to a synthesized, stable MOF structure. Most hypothetical MOFs generated through machine learning approaches have yet to be successfully synthesized as stable materials. Our approach aims to bridge this gap by providing more accurate binding energy predictions that can better guide experimental synthesis efforts and improve the practical relevance of computational screening

\section{Dataset Selection}
\label{sec:dataset}

Our study focusses on \acrshortpl{mof} from the MOF-74 family \cite{and2014a, ding2019a}. The MOF-74 series (also known as M2(dobdc) or CPO-27-M, where M represents different metal centers) is particularly interesting due to its high density of open metal sites, which leads to strong interactions with \coo molecules. The honeycomb-like structure features one-dimensional channels with a diameter of about 11 Å, making it ideal for gas separation applications. This family has the common linker \acrfull{dodbc} and has been well studied in the context of \coo capture meaning that experimental data for adsorption is readily available\cite{and2014a, ding2019a}. This does mean that aspects of this workflow are specific for this linker. However, throughout this section we will highlight how the process would be augmented when considering a wider set of families in a larger study, in particular some of the implications for how automated this workflow could be made.

The structures used in this study were taken from the Computation-Ready, Experimental Metal–Organic Framework (CoRE-MOF) database \cite{and2019b}. The selection of structures was based on the availability of reliable experimental data and the presence of well-characterized metal centers. The CoRE-MOF database is particularly valuable as it provides computationally-ready crystallographic information that has been cleaned and validated for simulation studies. However, it has been recently reported in the literature that the CoRE-MOF database contains a large number of MOFs with structural errors \cite{CoRE_shortcomings}. To minimize the risk of including a corrupted structure, the following steps were followed: First, we only considered structures associated to an experimental crystal structure from the Cambridge Structural Database (CSD), which allowed us to compare and validate the CoR-MOF structure; second, we prioritized structures that have been included in several other curated datasets, decreasing the likelihood of having unnoticed errors; and third, we inspected the structures looking for any potential structural error. 

 The selected structures are reported in Table \ref{table:mofs}, which shows the amount of \coo and heat of adsorption at a pressure of 1 bar and $25^{\circ}\text{C}$ temperature, where values were taken from \cite{choe2021a}. The selected MOFs represent a systematic variation of the metal center (Co, Fe, Ni, Cu, and Zn) while maintaining the same organic linker framework. This allows us to isolate and study the specific impact of metal substitution on \coo adsorption properties. The significant variation in CO2 adsorption capacity (from 2.9 to 7.1 $\text{mmol}\text{g}^{1}$) and heat of adsorption (from 5.3 to 9.3 kcal $\text{mol}^{-1}$) across the series demonstrates the crucial role of metal centers in determining adsorption behavior.

\begin{table}[h!]    
    \centering
    \begin{tabularx}{\linewidth}{@{}YYYYY@{}}
        \toprule
        MOF                         & CoRE-MOF ASR ID           & \begin{tabular}{@{}c@{}} $\text{CO}_{2}$ adsorption\\(mmol $\text{g}^{-1}$) \end{tabular} & \begin{tabular}{@{}c@{}}$Q_{s}$ \\ (kcal $\text{mol}^{-1}$) \end{tabular} \\
        \midrule
        $\text{Co}_{2}(\text{dobdc})$ & KOSKIO\_clean             & 6.9            & 8.1 \\
        $\text{Fe}_{2}(\text{dobdc})$ & ja205976v\_si\_006\_clean & 7.1            & 7.9 \\
        $\text{Ni}_{2}(\text{dobdc})$ & ORIVUI\_clean             & 7.0            & 9.3\\
        $\text{Cu}_{2}(\text{dobdc})$ & COKMOG\_clean             & 2.9            & 5.3 \\
        $\text{Zn}_{2}(\text{dobdc})$ & COKNAT\_clean             & 5.9            & 6.5 \\
    \end{tabularx}
    \caption{List of structures computed in the current study as identified by the CoRE-MOF ASR ID databases 
    IDs and its components in terms of metal center and organic linker. (dobdc stands for 2,5-dihydroxy-1,4-
    benzenedicarboxylic acid) alongside experimental values for \coo absorption and heat of adsorption taken 
    from \cite{choe2021a}. The groundstate spin per metal site is also show in which Co - Cu are based on ferromagnetic coupling between the metal centers and Zn is based on diamagnetic metal centers.}
    \label{table:mofs}
\end{table}

\section{Hierarchical Cluster Model}
\label{sec:clusters}

The simulations in this work are applied using a localized cluster model in which a finite section is cut out
from the \acrshortpl{mof} bulk, reducing a problem for an infinite solid to a finite fragment. Since the
stabilization of an absorbed \coo molecule is controlled by the local environment around the coordination
site, building clusters with centers at the coordination site is a sensible choice.
To calculate the binding energy between a \acrshort{mof} structure and a single \coo molecule we perform 3
groundstate calculations. First, we calculate the groundstate energy of the isolated \acrshort{mof} and \coo,
denoted $E_{\text{MOF}}$ and $E_{\text{CO}_{2}}$ respectfully as well as the groundstate energy when a single
\coo is adsorbed, denoted $E_{\text{MOF}+\text{CO}_{2}}$. With this the binding energy can be calculated as:

\begin{equation}
\label{eq:1}
    \Delta E = E_{\text{MOF}+\text{CO}_{2}} - (E_{\text{MOF}} + E_{\text{CO}_{2}}) 
\end{equation}

\begin{figure}[htp]
    \centering
    \includegraphics[width=18cm]{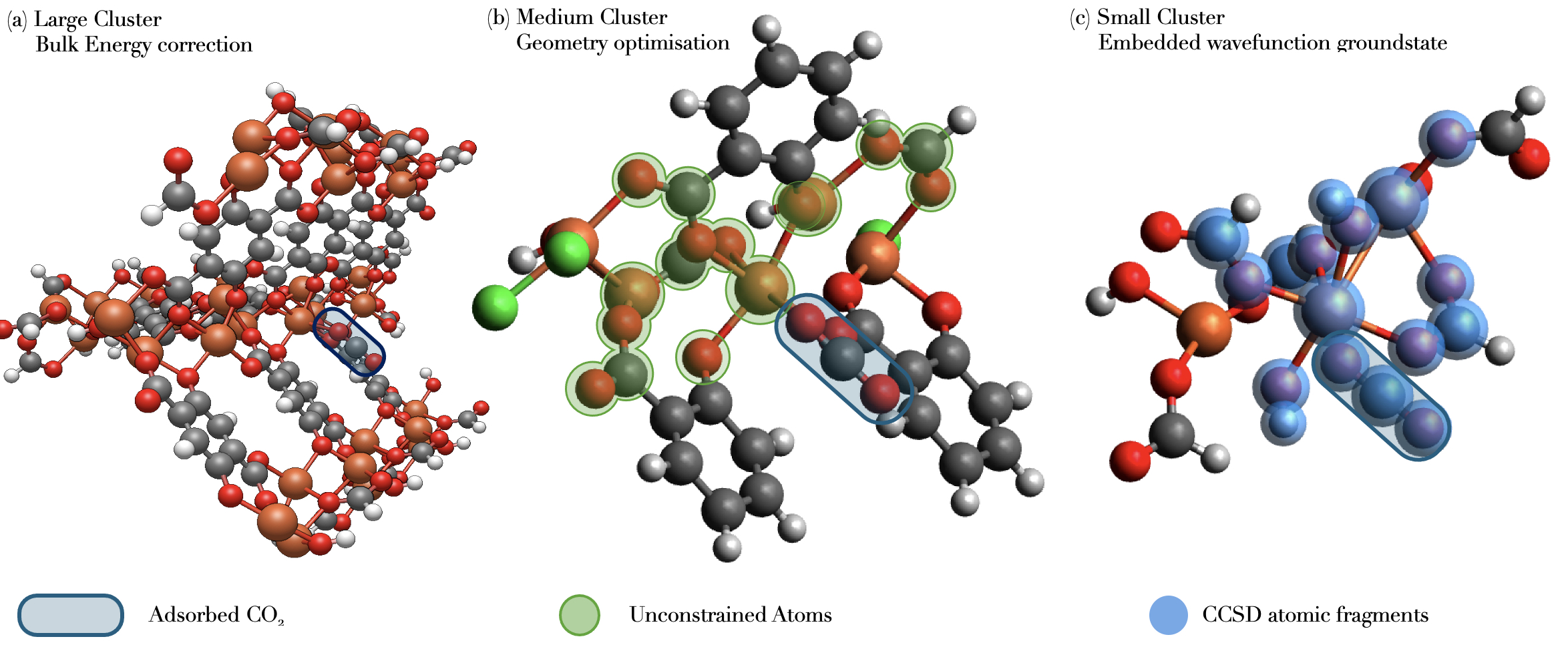}
    \caption{Overview of our multi-cluster approach shown for $\text{Fe}_2(\text{dobdc})$. (a) A large cluster model constructed by including all atoms within a certain radius (12.5 \r{A}). used to include the effects of the bulk material into binding energy calculations - see section (b) A medium cluster model that is used to perform constrained geometry relaxation for the large and small clusters. (c) A small cluster model where we apply embedded wavefunction calculations by splitting the cluster into atomic fragments. Here the atoms circled in red are solved using equation \ref{eq:3} that includes both a CCSD and MP2 calculation while the remaining atoms only use the MP2 calculation.}
    \label{fig:clusters}
\end{figure}

Each MOF cluster is formed by constructing a supercell from the corresponding structure file listed in column 2 of Table \ref{table:mofs}. We then perform a semi-automated process to include all atoms within a certain radius of the metal centre chosen to coordinate the \coo molecule. To handle linkers that cross the radius boundary we do the following; If all the linker’s non-hydrogen atoms are within the radius, the whole linker is kept otherwise only the carboxylate and alkoxide groups attached to a metal of the generated cluster are kept which are then modelled by either formate or hydroxylate respectfully. When considering applying this to different MOF families the choice of models to replace linkers that cross the boundary must be defined and thus this modelling step as well as the physical radius are the first set of parameters needed for this workflow. For the structures considered in this work we set the large cluster radius to 12.5 \r{A} (see Figure \ref{fig:clusters}(a) for $\text{Fe}_2(\text{dobdc})$).

We must acknowledge here that the use of a unit cell and periodic boundary conditions (PBC) on lieu of the larger cluster could be a more computationally efficient approach. However, bearing in mind that PBC calculations will provide no advantage for large systems lacking crystal symmetry, we adopted a more general cluster approach that is equally suited for periodic systems as for systems lacking periodicity.  

A smaller cluster, embedded within the large cluster, is defined (see Figure 1c), so higher-level, more accurate methodologies can be applied to it. We combine this accurate calculation on the small cluster with a lower accuracy calculation on the rest of the system using an ONIOM subtractive scheme [18]. Here, the energy of the larger cluster at a given high-level ($E^{HL}_{large}$), can be approximated by

\begin{equation}
\label{eq:2}
    E^{\text{HL}}_{\text{large}} \simeq E^{\text{HL}}_{\text{small}} + [E^{\text{LL}}_{\text{large}} - E^{\text{LL}}_{\text{small}}] 
\end{equation}

in which the HL/LL surperscripts indicate if the calculation is performed using the high- or low-level, respectively, and the large/small subscripts indicate if the calculation is performed on the large or the small cluster. So, the last term of equation \ref{eq:2} allows the effect of the bulk material on the MOF-\coo binding energies to be captured approximately at the low-level, while applying the high level to the binding center and its close environment. 

In this work, the small cluster is constructed by selecting the three metal centres closest to the adsorbed \coo and all its linkers. The linkers are however modeled by hydroxyl/formate moieties, which are the same models used to deal with linkers that cross the large cluster radius boundary, to keep the cluster at a reasonable size.

We use ferromagnetic states for all groundstate energy calculations except for $\text{Zn}_2(\text{dobdc})$ and $\text{Mg}_2(\text{dobdc})$ which are based on diamagnetic metal centres. This choice is based on comparing the energy of different spin eigenstates for the small cluster using DFT with the \acrshort{m06l} functional [19] and the def2-SVP basis set [20]. These calculations show that all MOFs except Cu and Ni favour the ferromagnetic state - in which the spins of neighbouring metal centres in the MOFs are coupled in a parallel fashion. For Cu and Ni, ferromagnetic and anti-ferromagnetic states were found to be very close in energy. However, for the purpose of calculating binding energies, we expect that being high or low-spin be a more important factor than the magnetic coupling between metallic centres. Practically, for our study this means that groundstate calculations are performed with a spin of 3 (Co), 4 (Fe), 2 (Ni), 1 (Cu), 0 (Zn and Mg) times the number of metal atoms in each cluster.

For the binding calculations we also need to add the adsorbed \coo molecule to the binding site proximal to the metal binding centre of the cluster. To optimize the position of the \coo, it is unpractical to use the large cluster due to its size and performing the optimization on the small cluster would neglect physical, important interactions of the \coo with atoms not included in the small cluster. For this reason we introduce an intermediate cluster soley for the purpose of geometry relaxation that is between the size of the small and large cluster size. This intermediate cluster is shown for $\text{Fe}_2(\text{dobdc})$ in Figure \ref{fig:clusters}(b).

In this work, the intermediate cluster contains five metal centers, and the organic linkers have been heavily modeled according to the following 3 criteria:

\begin{enumerate}
  \item For every linker whose phenyl ring is included in the cluster, only the carboxylate and alkoxide group directly coordinated to one of the included metals are preserved. The other carboxylate and alkoxide groups of the linker are removed and the ring is capped with a hydrogen linker atom.
  \item For organic linkers whose phenyl ring is not part of the cluster, the carboxylate and alkoxide groups attached to the considered metal centres are preserved and modelled by formate and hydroxyl anions, respectively. Each of those anionic moieties includes a hydrogen atom linker, capping the ligand in the position where the corresponding C-C or O-C bond was cleaved.
  \item We also included some selected Cl- ligands (green balls in Figure \ref{fig:clusters}(b)) to complete the coordination sphere of the metals at the edge of the cluster.
\end{enumerate}

To preserve the constraints of the bulk material only a subset of atoms was allowed to move during the geometry optimization, including the three central metals that are kept as part of the small cluster and all hydrogens that do not correspond to any capping atoms added to create the medium cluster. In Figure \ref{fig:clusters}(b) the atoms allowed to move during geometry relaxation are highlighted in green for $\text{Fe}_2(\text{dobdc})$. The geometry optimizations on the medium cluster were performed using the M06L functional together with a def2-SV\_P basis set as implemented in the Psi4 Python library \cite{and2020b}. The geometry relaxation was performed with and without the adsorbed \coo, and then the coordinates of the corresponding atoms were propagated to both the large and small clusters for consistency.

\section{Small Cluster Quantum Embedding Method}
\label{sec:embedding}

In this section we consider the steps to calculate the high accuracy groundstate of the small cluster, Esmallhigh in Equation 2. For this calculation we apply the EWF quantum embedding method as implemented in Vayesta python package \cite{nusspickel-a}. For a detailed account of this method, we refer the reader to \cite{nusspickel2022a} and \cite{nusspickel2023a}. To apply this method, we partition the full small cluster in into atomic fragments defined by the intrinsic atomic orbital (IAO) basis of each atom. While different fragmentation strategies are possible splitting the system at the atom level ensures that we have a consistent fragmentation strategy that can be applied when the adsorbed \coo is included or not and when we are comparing MOFs that use different linkers.

The coupling of the fragment IAO orbitals is described by a set of fragment bath orbitals. In the EWF method the bath is expanded beyond the traditional DMET \cite{knizia2013a} bath orbitals using and additional set of bath natural orbitals (BNO), which can be systematically enlarged to describe the beyond-mean-field coupling of the fragment to the environment at the level of the solver applied to the fragment \cite{nusspickel2022a}. The completeness of this bath space is controlled by a threshold $\eta$, where the bath becomes complete as $\eta \rightarrow 0$ and limited to just the traditional DMET bath as $\eta$ becomes large.

The IAOs of an atomic fragment combined with the bath orbitals are referred to as the \textit{fragment-cluster}. In \cite{nusspickel2023a} they outline several ways in which the groundstate solutions for each fragment cluster’s effective Hamiltonian can be combined to construct global expectation values over the whole system from the traditional DMET approach involving the partitioning of fragment cluster reduced density matrices or cumulants to partitioning of the fragment cluster wavefunctions. In this work we use the latter ”Global Expectation Values from Cluster Wave Functions” method to combine the fragment cluster solutions to calculate the groundstate energy of the whole small cluster, $E^{HL}_{small}$.

One of the features we can exploit in this method is that both the bath size, controlled by $\eta$ and the choice of quantum solver used can be different for each atomic fragment. We exploit this by using a less costly solver, MP2 with a large bath size, $\eta = 1 \times 10^{-7}$ combined with a more costly solver CCSD with a smaller bath size of $\eta = \eta_{\text{CCSD}} = 1 \times 10^{-5}$. The calculation for each atomic fragment is thus approximated using a multi-level approach \cite{cramer2004a, jensen2017a} utilizing MP2 and CCSD solvers, with different degrees of fragment-bath coupling. This can be written as:

\begin{equation}
\label{eq:3}
    E^{\text{CCSD}}_{\text{small}} \simeq E^{\text{CCSD}}_{\text{small}}(\eta_{\text{CCSD}}) + \left[E^{\text{MP2}}_{\text{small}}(\eta)-E^{\text{MP2}}_{\text{small}}(\eta_{\text{CCSD}})\right] 
\end{equation}

Where $E^{\text{CCSD/MP2}}_{\text{small}}(\eta)$ is the correlation energy found when applying the EWF approach to the small cluster for a given \textit{Solver} (CCSD or MP2, in this study) and a given value of $\eta$, and $E^{HF}_{small}$ is the reference mean-field energy for the small cluster calculations for which we use an unrestricted HF solver as implemented in pySCF \cite{and2020a}.

We can further increase the computational efficiency of this approximation by restricting the CCSD calculations to a subset of atomic fragments close to the metal binding site of the adsorbed \coo. In this work, we included the central metal binding site and then any additional atom that can be connected by at most 2 bonds, (and the \coo moiety, if present in the structure). In Figure \ref{fig:clusters}(c) we show these close atoms highlighted as blue spheres for $\text{Fe}_2(\text{dobdc})$. This modifies the energy in Equation \ref{eq:3} so that the correlation energies computed using $\eta = \eta_{CCSD}$ (namely, $E^{\text{CCSD}}_{\text{small}}(\eta_{\text{CCSD}})$ and $E^{\text{MP2}}_{\text{small}}(\eta_{\text{CCSD}})$) are restricted to the atomic fragments defined as close.

\section{Results and Discussion}
\label{sec:results}

In this section, we will provide the main results of the study, which is a comparison of the binding energy calculated following the steps in sections \ref{sec:clusters} - \ref{sec:embedding} with the experimentally measured heat of adsorption shown in Table \ref{table:ewf_energetics}.

\begin{table}[h!]    
    \centering
    \begin{tabularx}{\linewidth}{@{}YYYYY@{}}
        \toprule
        MOF                         & \begin{tabular}{@{}c@{}} $\Delta E$ (UHF)\\($\text{kcal}\,\text{mol}^{-1}$) \end{tabular}               & \begin{tabular}{@{}c@{}} $\Delta E$ ($\eta_{\text{CCSD}} = 1 \times 10^{-2}$)\\($\text{kcal}\,\text{mol}^{-1}$) \end{tabular} & \begin{tabular}{@{}c@{}} $\Delta E$ ($\eta_{\text{CCSD}} = 1 \times 10^{-5}$)\\($\text{kcal}\,\text{mol}^{-1}$) \end{tabular} \\
        \midrule
         $\text{Co}_{2}(\text{dobdc})$ & -3.417   & -5.586 & -6.129  \\
         $\text{Fe}_{2}(\text{dobdc})$ & -5.342   &  -6.754 & -7.454 \\
         $\text{Ni}_{2}(\text{dobdc})$ & -0.688   &  -3.850 & -4.655 \\
         $\text{Cu}_{2}(\text{dobdc})$ & -4.328   &  -6.069 & -6.281 \\
         $\text{Zn}_{2}(\text{dobdc})$ & -3.564   &  -2.496 & -4.578 \\
        \midrule
        $|\overline{\Delta E - Q_s}|$	& 4.549	 & 2.782	& 1.999
    \end{tabularx}
    \caption{Calculated binding energies using equations \ref{eq:1}, \ref{eq:2} 
    and \ref{eq:3}. For all results, the large and small cluster low accuracy energies were computed using M06L/def2-SV\_P as low-level. The triple-$\zeta$ def2-TZVP 
    basis set was used in combination with the EWF solvers to compute the small
    cluster energy contributions, see equation \ref{eq:3}. For the left column 
    only, the UHF/def2-TZVP was used as high-level method, while for the middle 
    and right column the EWF quantum method using CCSD bath size of $\eta_{\text{CCSD}} = 1 \times 10^{-2} \text{and} 1 \times 10^{-5}$, 
    respectively. In the bottom row we show the mean value of $|\Delta E-Q_s|$, where $Q_s$ is the heat of adsorption taken from Table \ref{table:mofs}, is shown for comparison.}
    \label{table:ewf_energetics}
\end{table}

\begin{figure}[htp]
    \centering
    \includegraphics[width=17cm]{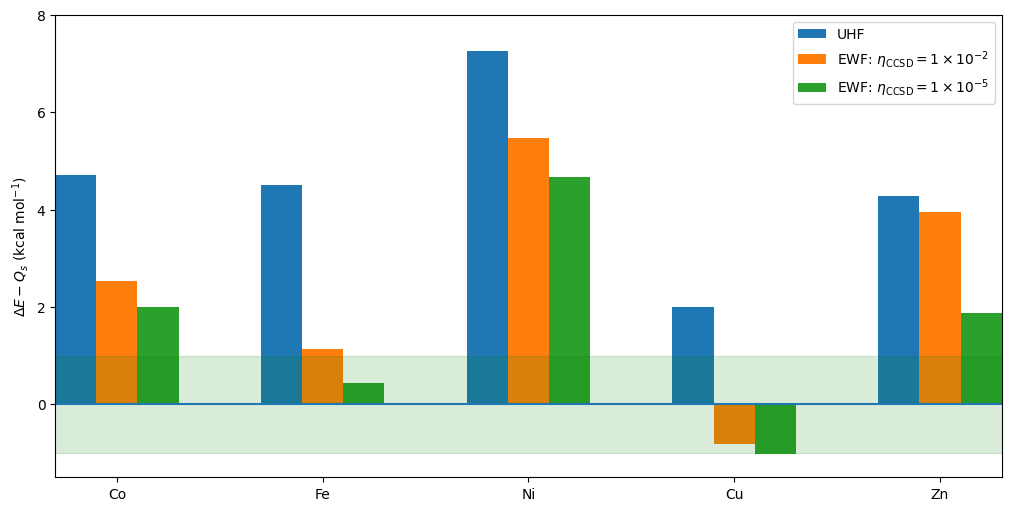}
    \caption{Plot of the $|\Delta E-Q_s|$ using equation \ref{eq:1} and the values from Table \ref{table:ewf_energetics} calculated using the hierarchical cluster model method outlined in sections \ref{sec:clusters} - \ref{sec:embedding}. For all values, the the low level corresponds to M06L/def2-SV\_P. The triple-$\zeta$ def2-TZVP basis set was used in combination with the EWF solvers to compute the small cluster energy contributions, see equation \ref{eq:3}. The blue bars show the energy when the UHF/def2-TZVP is used as high-level method, while the orange (green) bars shown uses the EWF method with a CCSD bath sizes of $\eta_{\text{CCSD}} = 1 \times 10^{-2}$ and $1 \times 10^{-5}$, see equation \ref{eq:3}.}
    \label{fig:energy_ewf}
\end{figure}

\begin{figure}[htp]
    \centering
    \includegraphics[width=17cm]{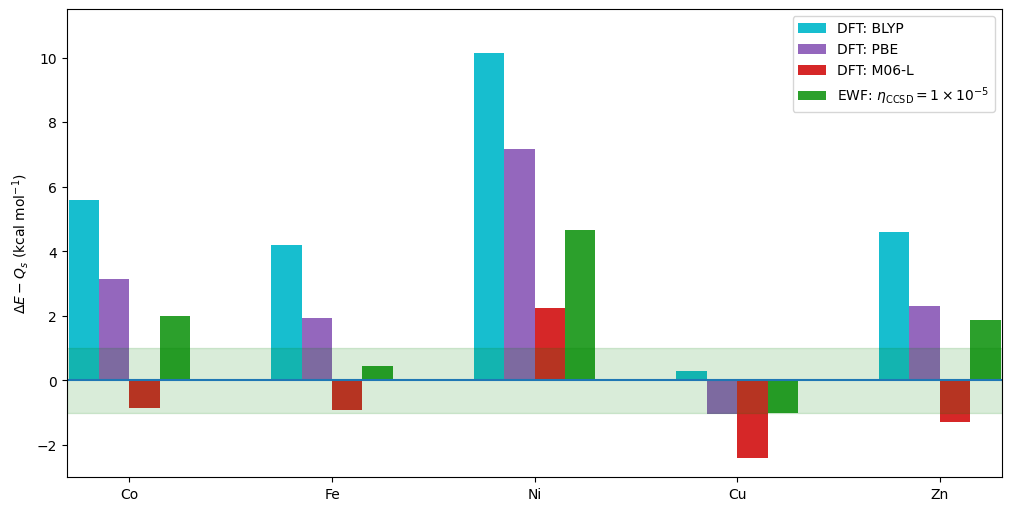}
    \caption{Plot of the $|\Delta E - Q_s|$ using equation \ref{eq:1} and the values from Tables \ref{table:ewf_energetics} and  \ref{table:dft_energetics} calculated using the hierachical cluster model method outlined in sections \ref{sec:clusters} - \ref{sec:embedding}. For all values, the the low level corresponds to M06L/def2-SV\_P. The triple-$\zeta$ def2-TZVP basis set was used in combination with three DFT functionals (BLYP, PBE, M06L) and the EWF method to compute the small cluster energy contributions, see equation \ref{eq:3}. The crossed values use DFT with functionals BYLP (blue), PBE (orange) and MO6L (black) The green dots shows results the EWF method with a CCSD bath size of $1 \times 10^{-5}$, see equation \ref{eq:3}.}
    \label{fig:energy_dft}
\end{figure}

Tables \ref{table:ewf_energetics} and \ref{table:dft_energetics} show the ONIOM results computed using M06L/def2-SV\_P as low-level in all cases, in combination with a number of different high-level methodologies: UHF, EWF ($\eta_{CCSD} = 1 \times 10^{-2}$ and $\eta_{CCSD} = 1 \times 10^{-5}$) and DFT (M06L, BLYP, PBE). For all the high-level calculations, the def2-TZVP basis set was used. In Figure \ref{fig:energy_ewf} we compare the values of $\Delta E - \text{Q}_s$ for the ONIOM calculations employing UHF and the two EWF approaches herein considered. The plot shows a clear trend in which the results improve from Hartree Fock (blue dots) to the smaller bath size (orange dots) to the larger bath size (green dots). Here we show the boundary of chemical accuracy in light green ($\pm 1$ kcal $\text{mol}^{-1}$). This validates the idea that this approach provides a simple way to systematically improve the calculation accuracy, which is not possible for a given DFT functional. With that in mind, in Figure \ref{fig:energy_dft} we compare the results using three DFT functionals (BYPL, PBE and ML06) as high-level methods and include again the results using EWF as high-level method. EWF beats the BYLP and PBE methods. M06L results in a smaller value of the average error $\Delta E - \text{Q}_s = 1.536$ compared to $\Delta E - \text{Q}_s = 1.999$, a difference which is largely driven by the result for $\text{Ni}_2(\text{dobdc})$. There is, however, no systematic way to improve the DFT result for fixed basis. In contrast, the EWF method can use the choice of fragment cluster bath size and solver type to systematically improve the accuracy which was one of the key motivators for investigating this approach.
 
\begin{table}[h!]    
    \centering
    \begin{tabularx}{\linewidth}{@{}YYYYY@{}}
        \toprule
        MOF           & \begin{tabular}{@{}c@{}} $\Delta E$ (BLYP) \\($\text{kcal}\,\text{mol}^{-1}$) \end{tabular}               & \begin{tabular}{@{}c@{}} $\Delta E$ (PBE)\\($\text{kcal}\,\text{mol}^{-1}$) \end{tabular} & \begin{tabular}{@{}c@{}} $\Delta E$ (M06L)\\($\text{kcal}\,\text{mol}^{-1}$) \end{tabular} \\
        \midrule
         $\text{Co}_{2}(\text{dobdc})$ & -2.530   &  -4.978 & -8.979  \\
         $\text{Fe}_{2}(\text{dobdc})$ & -3.699   &  -5.962 & -8.7894 \\
         $\text{Ni}_{2}(\text{dobdc})$ & -0.816   &  -2.147 & -7.085 \\
         $\text{Cu}_{2}(\text{dobdc})$ & -4.962   &  -6.305 & -7.667 \\
         $\text{Zn}_{2}(\text{dobdc})$ & -1.844   &  -4.143 & -7.732 \\
        \midrule
        $|\overline{\Delta E - Q_s}|$	& 4.9652 & 3.1208 & 1.536
    \end{tabularx}
    \caption{Calculated binding energies using equations \ref{eq:1}, \ref{eq:2} and \ref{eq:3}. For all results the large and small cluster low accuracy energies, using M06L/def2-SV\_P as low-level. Three different DFT functionals where used as high-level method in combination with a triple-$\zeta$ def2-TZVP basis set. In the bottom row we show the mean value of $|\Delta E - Q_s|$, where $Q_s$ is the heat of adsorption taken from Table \ref{table:mofs}, is shown for comparison.}
    \label{table:dft_energetics}
\end{table}

When comparing our computational binding energies with experimental heat of adsorption measurements, we must acknowledge inherent methodological differences that affect direct comparisons. Our calculations represent binding energies at zero temperature and infinite dilution (single molecule adsorption), while experimental measurements are conducted at finite temperature (25°C) and non-zero gas loading conditions. This introduces systematic deviations due to temperature effects (entropic contributions absent in ground state calculations), loading effects (adsorbate-adsorbate interactions at experimental conditions), and framework flexibility considerations. Despite these limitations, the relative trends across different MOF structures remain informative for screening purposes, and our systematic approach allows for consistent comparisons between candidate materials. Future work could incorporate molecular dynamics simulations and finite temperature effects to further bridge this gap between computational models and experimental conditions.

\section{Summary}
\label{sec:summary}

In this work, we outline a hierarchical cluster model that combines larger scale DFT calculations with a quantum embedding method that allows for better systematic convergence of accuracy. This approach addresses a critical challenge in computational materials science: balancing computational efficiency with chemical accuracy when evaluating complex porous materials such as metal-organic frameworks for carbon capture applications.

Our methodology offers several distinct advantages over traditional computational approaches. First, the hierarchical cluster model effectively captures the local chemical environment of the \coo binding site while accounting for the extended framework effects, providing a more realistic representation of the adsorption process. Second, the embedded wavefunction approach allows for systematic improvement of accuracy through bath size control, offering a clear pathway to convergence that is not available in standard DFT implementations. Third, our multi-level approach efficiently allocates computational resources by applying higher-level methods only to the most critical regions of the system.

A key innovation in our approach is the use of wavefunction-based embedding rather than density matrix-based methods. By directly projecting wavefunction amplitudes instead of reduced density matrices, we achieve more rapid convergence with respect to bath size and more accurate treatment of strong correlation effects. This distinction becomes particularly important when considering the integration with quantum computing hardware, where extracting full wavefunction information presents different challenges than obtaining reduced density matrices. Our framework thus provides a foundation for future quantum-classical hybrid approaches that can leverage emerging quantum advantage while maintaining computational tractability.

The validation against experimental heat of adsorption data for the MOF-74 series demonstrates the potential of this approach for high-throughput screening applications. While computational screening of MOFs has generated thousands of hypothetical structures, the translation to experimentally viable materials remains challenging. Our method's improved accuracy in predicting binding affinities could help bridge this gap by providing more reliable guidance for experimental synthesis efforts.

We acknowledge the inherent limitations in comparing our zero-temperature, infinite-dilution calculations with finite-temperature experimental measurements. Nevertheless, the systematic nature of our approach ensures consistent relative comparisons between candidate materials, which is valuable for screening purposes. Future work will focus on incorporating temperature and loading effects to further enhance the predictive power of our computational framework.

The workflow presented here is designed with scalability in mind, allowing for extension beyond the MOF-74 family to explore the vast chemical space of potential MOF structures. Additionally, the quantum embedding framework provides a natural entry point for quantum computing algorithms, potentially enabling even more accurate predictions as quantum hardware capabilities advance. By combining classical and quantum computational approaches, this work represents a step toward more efficient discovery and optimization of materials for carbon capture and other critical environmental applications.

Furthermore, our approach addresses several persistent challenges in the computational modeling of MOFs. The heterogeneous nature of metal-organic binding sites often requires treatment beyond standard DFT approximations, particularly when transition metals with complex electronic structures are involved. Our embedding method provides a more rigorous treatment of these electronic correlation effects while maintaining computational tractability. The semi-automated workflow we've developed also reduces the human intervention required for setting up complex calculations, making it feasible to extend this approach to diverse MOF families with varying metal centers and organic linkers.

The implications of this work extend beyond carbon capture applications. The systematic accuracy control demonstrated here could benefit computational studies of other gas separation processes, catalytic reactions at MOF nodes, and the design of materials for energy storage applications. As computational resources continue to advance and quantum computing capabilities mature, the framework established in this study provides a foundation for increasingly accurate in silico materials design, potentially accelerating the discovery cycle for functional materials addressing urgent environmental and energy challenges.

\section{Perspectives of using quantum hardware} \label{sec:qhardware}

Like other quantum embedding methods, EWF provides a promising pathway to apply groundstate algorithms executed on quantum hardware to larger chemical systems. For the workflow described in this work, we investigated replacing some of the CCSD calculations with hybrid variational quantum algorithms implemented on IBM QPUs. For the basis sets considered, this required an additional reduction of the atomic fragment clusters' orbitals via an active space transformation to ensure systems were sufficiently small for meaningful results from current hardware, even with advanced error mitigation methods applied using the Qiskit runtime \cite{qiskit}. Unfortunately, this reduction resulted in significant loss of correlation energy compared to using the complete atomic fragment cluster, making direct comparison with classical solvers challenging at present.

A key consideration specific to our wavefunction-based approach is that the groundstate calculation requires extracting a subset of wavefunction amplitudes from each fragment cluster solution, rather than just reduced density matrices as in methods like DMET. For quantum states implemented as quantum circuits, directly extracting wavefunction amplitudes presents significant technical challenges. While the EWF method can utilize reduced one- and two-body density matrices, this approach sacrifices some of the key advantages of wavefunction-based methods, particularly the smooth convergence of accuracy with bath size as demonstrated in \cite{nusspickel2023a}.

Recently a method that utilises quantum computers for groundstate calculations in conjunction with classical algorithms removes the need to extract wavefunction amplitudes directly from quantum hardware and is therefore a useful method that we will explore in future research. The method is called \acrfull{sqd} and has recently been implemented using IBM quantum hardware \cite{sqd} and also within a DMET framework \cite{knizia2013a}. One main point to mention about this method is that since this method is hybrid in nature, a selected CI solver stores the wavefunction amplitudes classically meaning this method can be efficiently integrated with our workflow without modification. While this offers opportunity to get the best out of nearer term quantum computers we believe that it is more important to understand how more general quantum solvers can be integrated with this workflow by making the extraction of the wavefunction amplitudes from states represented as quantum circuits as efficiently as possible and this is the focus of on-going research.

As quantum hardware continues to advance in both qubit count and coherence time, the ability to handle larger active spaces will improve, potentially enabling more complete fragment cluster calculations without the compromises currently necessary. Our ongoing research focuses on optimizing the interface between quantum embedding methods and quantum hardware, particularly addressing the challenge of efficiently extracting wavefunction information while preserving the systematic improvability that makes our approach valuable for materials discovery.

\section{Outlook: From Quantum Calculations to Carbon Capture Solutions}

The computational framework presented in this study represents a significant advancement toward accelerating the discovery and optimization of Metal-Organic Frameworks for carbon capture applications. While experimental testing of MOFs for CO$_2$ capture remains resource-intensive and time-consuming, our systematically improvable quantum embedding approach offers several strategic advantages for addressing global carbon capture challenges at scale.

The demonstrated accuracy of our calculations in matching experimental heat of adsorption data, particularly when using larger bath sizes, suggests this method could reliably predict CO$_2$ capture performance of novel MOF structures before synthesis. This predictive capability could substantially reduce the experimental search space, focusing synthetic efforts on the most promising candidates. Furthermore, the hierarchical cluster model, combined with the ability to tune computational accuracy through bath size parameters, makes this approach suitable for high-throughput screening of large MOF databases containing thousands of structures. This scalability is crucial for exploring the vast chemical space of potential MOF materials efficiently.

Our approach provides detailed insights into the local chemical environment affecting CO$_2$ adsorption at the electronic structure level, revealing how subtle variations in metal centers and coordination environments influence binding energies. These mechanistic insights could guide rational design strategies for next-generation MOF structures with enhanced selectivity, capacity, and regeneration properties—all critical factors for practical carbon capture technologies. The systematic nature of our calculations also enables reliable comparison between different MOF families, helping to identify promising structural motifs that could be incorporated into novel materials.

The framework developed here is well-positioned to leverage emerging quantum computing capabilities, potentially enabling even more accurate predictions for complex MOF systems as quantum hardware advances. The wavefunction-based embedding approach provides a natural integration point for quantum algorithms specifically designed to tackle strongly correlated electronic systems, which are often challenging for conventional computational methods. This quantum-classical hybrid approach could eventually enable accurate simulations of larger and more complex MOF structures than currently possible.

Beyond carbon capture, the methodology developed here could be extended to other critical environmental applications, including hydrogen storage, methane separation, and catalytic conversion of captured CO$_2$ into value-added products. The ability to systematically control accuracy while maintaining computational efficiency makes this approach valuable for any application where the precise energetics of gas-material interactions determine performance.

By providing a reliable computational pathway to evaluate MOF performance before synthesis, this work contributes to the broader effort of developing more efficient carbon capture technologies to address climate change challenges. The ability to computationally screen and optimize MOF structures could significantly reduce the time from concept to deployment for next-generation carbon capture materials, accelerating progress toward meeting global carbon reduction targets.

\section{Acknowledgments}
This work was supported by the Defense Advanced Research Projects Agency (DARPA) Accelerated Research in Circuits (ARC) program under the Innovating Mechanisms for Programmable and Automated Quantum Technologies (IMPAQT) initiative [HR0011-24-3-0002]. The authors are thankful to Juan Manuel Aguiar Hualde for insightful discussions surrounding embedding methods, which was crucial to laying the foundations of this study. The authors are also thankful to Yann Magnin from TotalEnergies for extensive and thorough feedback that significantly improved the quality of this manuscript.

\bibliographystyle{achemso}
\bibliography{main}
\end{document}